\documentclass[12pt]{article}
\begin{document}
\large
\par
\noindent {\bf Computation of the number of neutrino events which
can be registered in Borexino detector from the Sun neutrinos flux
with energy $E_\nu = 0.862 MeV$}

\par
\begin{center}
\vspace{0.3cm} Beshtoev Kh. M. (beshtoev@cv.jinr.ru)
\par
\vspace{0.3cm} Joint Institute for Nuclear Research, Laboratory of
High Energy Physics, Joliot Curie 6, 141980 Dubna, Moscow region,
Russia.
\end{center}
\vspace{0.3cm}

\par
{\bf Abstract}

\par
This paper gives an estimation of the number of neutrinos which
can be registered in Borexino detector from the Sun neutrinos
generated in reaction $^{7}Be + e^{-} \to ^{7}Li + \nu_e$ with
energy $E_{\nu_e} = 0.862 MeV$ in the absence of neutrino
oscillations. This number is supposed  be between $ N^{theor} =
86.45 \div 96.52 \frac{counters}{(day\cdot 100\quad ton)}$ in
dependence on primary neutrino fluxes. Then ratios between number
of neutrinos $N^{exper}$ registered in Borexino detector and
counted numbers $N^{theor}$, are $\frac{N^{exper}}{N^{theor}} =
0.49 \div 0.54$. This value is close enough to the same value
obtained in $^{71}Ga - ^{71}Ge$ experiments in the close energy
regions. The value $\frac{N^{exper}}{N^{theor}}$ obtained at
supposition that $\theta_{1 3} \approx 0$ and absence of the
resonance effect approximately equal to $\simeq 0.67$ and it is
noticeably greater than the above value. Probably it means that
the supposition that $\theta_{1 3} \approx 0$ is not justified and
there can be a definite deposit of $\tau$ neutrinos.
\\


\section{Introduction}

In the present time the detector Borexino goes on operating [1, 2
]. One of the major tasks of this detector is to measure of the
Sun neutrinos flux with energy $E_\nu = 0.862 MeV$, appearing in
reaction $^{7}Be + e^{-} \to ^{7}Li + \nu_e$. Measurements of the
neutrino flux at this energy are very important since in this
energy region the deposit of the resonance mechanism of neutrino
oscillations is very small [3, 4] then if to suppose that the
angle mixing $\theta_{1 3} \approx 0$ only electron neutrino
vacuum oscillations can be observed in this case. Usually it is
supposed that the deficit of high energy Sun neutrinos caused by
the resonance mechanism of neutrino oscillations in the Sun
matter.
\par
The value of $sin^2 2\theta_{1 2}$ obtained in work [5] from the
reactor experiment (neutrino vacuum oscillations) is
$$
sin^2 (2 \theta_{1 2}) \simeq 0.83, \eqno(1)
$$
then the fraction (part) of electron neutrinos $P_{\nu_e}$ is
$$
P_{\nu_e} = 1 - \frac{1}{2} sin^2 (2 \theta_{1 2}) \simeq 0.615,
\eqno(2)
$$
and the remaining neutrinos are muon ones and a relative portion
$P_{\nu_\mu}$ of these neutrinos is
$$
P_{\nu_\mu} = \frac{1}{2} sin^2 (2 \theta_{1 2}) \simeq 0.385.
\eqno(3)
$$
If electron neutrinos are registered via the charged current, then
$P_{\nu_e}(W)$ must be equal to 0.615.  But if neutrinos are
registered via the charged and neutral currents (as it takes place
in Borexino experiment), then we must add the deposit of neutral
current from the electron and muon neutrinos then $P_{\nu_e}(Z^o)
\simeq 0.155$ (see the value obtained in SNO [6]) and
$$
P_{\nu_e}(W, Z^o) \simeq 0.615 + 0.155 = 0.770 . \eqno(4)
$$
\par
In Borexino detector neutrinos are registered via the neutral and
charged currents. If the primary neutrino flux is $N^o_e$ electron
neutrinos, then if there are no neutrino oscillations, then this
detector can register $n$ neutrinos, which is a sum of events
registered via neutral current $n^o$ and charged current
$n^{neutral} = 0.155 \cdot n^o$ (value $0.155$ is relative portion
of generated by neutral current), then
$$
n = n^o + n^{neutral} = (1 + 0.155) n^o . \eqno(5)
$$
\par
If $\theta_{1 3} \approx 0$ and there are electron neutrino
oscillations then via the charged current $n^{charged} = n^o
 P_{\nu_e}$ neutrinos can be registered and via the neutral current
$n^{neutral} = 0.155 \cdot n^o$ neutrinos (all electron and muon
neutrinos interact via neutral current) can be registered and the
sum of neutrinos which can be registered in Borexino detector is
$$
n^{osc} = n^{charged} + n^{neutral} = (P_{\nu_e} + 0.155) n^o .
\eqno(6)
$$
The ratio between $ n^{osc}$ and $n$ is
$$
\frac{n^{osc}}{n} = \frac{(P_{\nu_e} + 0.155)}{(1 + 0.155)} =
0.667  . \eqno(7)
$$
This value is the value which can be obtained in Borexino detector
if $\theta_{1 3} \approx 0$ and there are only oscillations of
electron neutrinos.
\par
In work [1] it was reported that Borexino detector must detect
about $55\quad counts/(day \cdot 100\quad ton)$ neutrino events at
the absence of the resonance effect, then the following work [2]
reported that this detector must detect $75\pm 4 counts/(day \cdot
100\quad ton)$ at the absence of neutrino oscillations.
\par
The purpose of this work is an independent estimation of number of
neutrino events which can be registered in Borexino detector from
the Sun neutrinos flux with energy $E_\nu = 0.862 MeV$ at the
absence of neutrino oscillations.
\par
From the experiments we know that the angle mixing of $\theta_{2
3} \simeq \pi/4\quad (45^o)$ [7, 8] and $\theta_{1 2} \simeq 32^o$
[5]. The author holds the point of view that since the above
(other) angle mixings are big, then there is no reason to suppose
that the third angle of mixing $\theta_{1 3}$ can be very small
(analysis situation with this supposition see in work [9]).

\section{Flux of the Sun neutrinos from $^{7}Be + e^{-} \to ^{7}Li +
\nu_e$ reaction computed in framework of the Standard Sun Model}

The discussion of the Standard Sun Model (SSM) was given by J.
Bahcall in [10]. The flux of the sun neutrinos from reaction
$^{7}Be + e^{-} \to ^{7}Li + \nu_e$ obtained in [10], is
$$
N^{theor}_{\nu_e} = 0.47 (1 \pm 0.15)\cdot 10^{10} cm^{-2} c^{-1}
. \eqno(8)
$$
Afterward the neutrino flux from this reaction was carried out [3,
11] and values of $N^{theor}_{\nu_e}$ was calculated as
$$
N^{theor 1}_{\nu_e} = 0.455 \cdot 10^{10} cm^{-2} c^{-1} ,
\eqno(9)
$$
and
$$
N^{theor 2}_{\nu_e} = 0.508 \cdot 10^{10} cm^{-2} c^{-1} .
\eqno(10)
$$
In our computations (estimations) we will use the above values for
$N^{theor}_{\nu_e}$

\section{$\nu_e + e^{-} \to \nu_e + e^{-}$ elastic scattering cross section}

The elastic scattering of electron neutrino on electron is
realized via $W$ (charge current) and $Z$ (neutral current) boson
exchanges. In literature they usually take differential cross
cross section and elastic cross section obtained in [12] (see also
ref. [13]). Then the expression for differential cross section
abtains the following form:
$$
\frac{d \sigma_{\nu_e e} (W, Z)}{d T} =  \frac{ 2 m_e G^2_F}{\pi}
[(\frac{1}{2} + \xi)^2 + \xi^2 (1 - \frac{T}{E_{\nu_e}})^2 -
(\frac{1}{2} + \xi) \xi \frac{m_e T}{E_{\nu_e}^2}], \eqno(11)
$$
and the expression for the elastic cross section is:
$$
\sigma_{\nu_e e} (W, Z) = \frac{G^2_F s}{\pi} [(\frac{1}{2} +
\xi)^2 + \frac{1}{3} \xi^2] , \eqno(12)
$$
where $G_F$ is Fermi constant, $\xi = sin^2 \theta_W$, $t = (k_1 -
k_2)^2 = (p_2 - p_1)^2$ (if electron is in rest, then $t = 2 m^2_e
+ 2 m_e E_{e 2}, $ $s = (k_1 + p_1)^2$ (if electron is in rest,
then $s = m^2_e + 2 m_e E_{\nu_e}$), $T = (E_{2e} - m_e)$ - is
kinetic energy of the scattered electron.
\par
The expression for cross sections taking into account radioactive
corrections, is given in [14]. We will not use these corrections
since the uncertainty in calculated flux of the Sun neutrinos
considerably exceed these corrections.
\par
The expression for elastic cross section (12) after substitution
the value of $s$ has the following form:
$$
\sigma_{\nu_e e} (W, Z) = \frac{G^2_F m^2_e}{\pi} (1 + 2
\frac{E_{\nu_e}}{m_e}) [(\frac{1}{2} + \xi)^2 + \frac{1}{3} \xi^2]
\simeq
$$
$$
\simeq \frac{G^2_F m^2_e}{\pi} 2 (\frac{E_{\nu_e}}{m_e})
[(\frac{1}{2} + \xi)^2 + \frac{1}{3} \xi^2] =
$$
$$
= 1.722 \cdot 10^{-44}{E_{\nu_e}(MeV)} [(\frac{1}{2} + \xi)^2 +
\frac{1}{3} \xi^2]\quad cm^2 . \eqno(13)
$$
It is necessary to remark that this expression for the cross
section is correct only at high energies when $E_{\nu_e} >> m_e$.
At low energies we must take the threshold effect into account,
and then $E_{\nu_e}$ must be change on
$$
E_{\nu_e} \to \frac{E_{\nu_e}}{(1 + \frac{m_e}{2 E_{\nu_e}})}  ,
$$
Then
$$
\sigma_{\nu_e e} (W, Z) = 1.722 \cdot 10^{-44}
\frac{E_{\nu_e}(MeV)}{(1 + \frac{m_e}{2 E_{\nu_e}})} [(\frac{1}{2}
+ \xi)^2 + \frac{1}{3} \xi^2]\quad cm^2 . \eqno(14)
$$
The average value for $\xi$ [15] is $0.232\div 0.234$. However, it
is necessary to remark that for the best fitting expression (12)
to the experimental data measured in [16, 17] for $\nu_e + e^{-}
\to \nu_e + e^{-}$ elastic cross section, this value must be
$sin^2 \theta_W = 0.248$. Therefore in our computations
(estimations) we will use this value for $\xi$. Then the value for
expression $[(\frac{1}{2} + \xi)^2 + \frac{1}{3} \xi^2]$ is
$0.581$.  At $E_{\nu_e} = 0.862 MeV$ $\sigma_{\nu_e e} (W, Z) =
0.665\cdot 10^{-44} cm^2$ (now it is not necessary to use
radioactive corrections since we have made normalization for the
cross section measured in experiments [16, 17]).
\par
In literature there is another expression for $\nu_e + e^{-} \to
\nu_e + e^{-}$ elastic scattering [18]. These expressions coincide
at high energies $E_{\nu_e}$, but they differ at low energies
$E_{\nu_e}$. Probably, it is necessary to find out the reason of
their discrepancies.

\section{Characteristics of liquid scintillator of Borexino Detector}

Borexino detector uses trimethylbenzene $C_6 H_3 (C H_3)_3$ (or
$C_9 H_{12}$) and as scintillator $C_{15} H_{11} N O \quad (1.5
\quad g/l)$ [1, 2] is used. The density of trimethylbenzene is
$\rho = 0.8761\quad g\cdot cm^{-3}$ (in our calculations we will
not take into account this small addition related with the
scintillator). The molecular weight of trimethylbenzene is $B =
120.19 \quad g\cdot mol^{-1}$. Then number of $C_6 H_3 (C H_3)_3$
molecules $n_M$ in one $cm^3$ is
$$
n_{M} = \frac{\rho}{B} N_A = 4.389 \cdot 10^{21}, \eqno(15)
$$
where $N_A$ is the Avogadro number.
\par
The one molecule of $C_6 H_3 (C H_3)_3$ includes 66 electrons,
then the number of electrons $n_e$ in $1 cm^3$ of
trimethyl\-benzene (i. e., electron density) is
$$
n_e = 2.897 \cdot 10^{23}\quad cm^{-3}, \eqno(16)
$$
then 100 ton ($G = 10^5 \quad g$) of  trimethylbenzene contains
$$
N_e = \frac{n_e G}{\rho} = 3.307 \cdot 10^{31} , \eqno(17)
$$
electrons.

\section{Estimation of the number of neutrinos which can be registered in Borexino detector
from the Sun neutrinos with $E_{\nu_e} = 0.862$ MeV}

Using the previous computations now we estimate the number of
neutrinos $N_B$ (event rates) which can be registered in the
Borexino detector in $100\quad ton$ of  trimethylbenzene during of
one day ($t = 8.64 \cdot 10^4 s$).
$$
N^{1 theor} = N_e \cdot \sigma_{\nu_e e}(W, Z, 0.862 MeV)\cdot t
\cdot N^{theor 1}_{\nu_e} = 86.45, \eqno(18)
$$
and
$$
N^{2 theor} = N_e \cdot \sigma_{\nu_e e}(W, Z, 0.862 MeV)\cdot t
\cdot N^{theor 2}_{\nu_e} = 96.52. \eqno(19)
$$
So, the above estimations have shown that the event rates on 100
ton/day in Borexino detector from the Sun neutrinos with
$E_{\nu_e} = 0.862$ MeV can be as follows:
$$
N^{theor} = 86.45 \div 96.52 \frac{counters}{(day\cdot 100\quad
ton)} . \eqno(20)
$$
In [2] it was reported that the rate of events registered in this
detector is
$$
N^{exper} = 47 \pm 7 (stat.) \pm 12(syst.)
\frac{counters}{(day\cdot 100\quad ton)} . \eqno(21)
$$
Thus, the portion of neutrinos $N^{exper}$ registered in this
experiment relative to the computations in the framework of SSM
are
$$
\frac{N^{exper}}{N^{theor}} = 0.49 \div 0.54  .   \eqno(22)
$$
This value is close enough to the same value obtained in $^{71}Ga
- ^{71}Ge$ experiments [19, 20] in the energy regions near to
above.
\par
The value for $\frac{N^{exper}}{N^{theor}}$ in expr.(22) is
noticeably smaller than the value (0.67) obtained in expr.(7) if
to suppose that $\theta_{1 3} \approx 0$. Probably it means that
this supposition is not justified and there can be a definite
deposit of $\tau$ neutrinos.

\section{Conclusion}

In this work the number of neutrinos which can be registered in
Borexino detector from the Sun neutrinos generated in reaction $
^{7}Be + e^{-} \to ^{7}Li + \nu_e$ with energy $E_{\nu_e} = 0.862
MeV$ was calculated. This number is between
$$
N^{theor} = 86.45 \div 96.52 \frac{counters}{(day\cdot 100\quad
ton)}, \eqno(23)
$$
in dependence on the primary neutrino fluxes. Then the ratios
between neutrinos $N^{exper}$ registered in this experiment and
the calculated numbers $N^{theor}$ are as follows
$$
\frac{N^{exper}}{N^{theor}} = 0.49 \div 0.54  .   \eqno(24)
$$
This value is close enough to the same value obtained in $^{71}Ga
- ^{71}Ge$ experiments [19, 20] in the close energy regions. The
value $\frac{N^{exper}}{N^{theor}}\simeq 0.67$ obtained in (7) at
supposition that $\theta_{1 3} \approx 0$ and the resonance effect
is absent is noticeably more than the value in (24). Probably, it
means that supposition that $\theta_{1 3} \approx 0$ is not
justified and there can be a definite deposit of $\tau$ neutrinos.
\\


\end{document}